# Influence of the Leaving Group on the Dynamics of a Gas Phase S$_N$2 Reaction

Martin Stei[1], Eduardo Carrascosa[1], Martin A. Kainz[1*], Aditya H. Kelkar[1†], Jennifer Meyer[1], István Szabó[2,3], Gábor Czakó[3], Roland Wester[1]

[1] Institute for Ion Physics and Applied Physics, Universität Innsbruck, Technikerstr. 25, 6020 Innsbruck, Austria

[2] Laboratory of Molecular Structure and Dynamics, Institute of Chemistry, Eötvös University, P.O. Box 32, 1518 Budapest 112, Hungary

[3] Department of Physical Chemistry and Materials Science, University of Szeged, Rerrich Béla tér 1, Szeged, H-6720, Hungary

**In addition to nucleophile and solvent, the leaving group has a significant influence on nucleophilic substitution (S$_N$2) reactions. Its role is frequently discussed with respect to reactivity, but its influence on the reaction dynamics remains obscured. Here, we uncover the influence of the leaving group on the gas phase dynamics of S$_N$2 reactions in a combined approach of crossed-beam imaging and dynamics simulations. We have studied the reaction F$^-$ + CH$_3$Cl and compared it to F$^-$ + CH$_3$I. For the two leaving groups Cl and I we find very similar structures and energetics, but the dynamics show qualitatively different features. Simple scaling of the leaving group mass does not explain these differences. Instead, the relevant impact parameters for the reaction mechanisms are found to be crucial, which is attributed to the relative orientation of the approaching reactants. This effect occurs on short time scales and may also prevail under solution phase conditions.**

---

[*] Present address: Photonics Institute, TU Wien, Gußhausstraße 27-29, 1040 Vienna, Austria
[†] Present address: Department of Physics, Indian Institute of Technology Kanpur, Kanpur-208016, Kanpur, India

Bimolecular nucleophilic substitution (**S$_N$2**) plays a pivotal role in chemical synthesis, especially for interchanging functional groups and for carbon-carbon bond formation[1]. Consequently, it is one of the most widely studied reactions in physical organic chemistry[2–16]. Usually, solvent effects are superimposed onto the intrinsic reaction dynamics, which strongly affects the reactivity. However, it is now becoming increasingly clear that on short time scales direct atomistic dynamics of chemical reactions can prevail - at least partially - in a solution phase environment[17–19]. By studying S$_N$2 reactions in the gas phase these intrinsic dynamics, mechanisms and structure-energy relations become accessible.

Gas phase S$_N$2 reactions are characterized by a double well potential energy surface[2], which stems from the intermediate ion-dipole complexes in the entrance and exit channels on either side of the central barrier. For many exothermic reactions the central barrier lies below the energy of reactants, nevertheless, it has a substantial influence on the reaction kinetics by introducing dynamical constraints with respect to angular momentum, intramodal energy transfer, or steric bottlenecks[20]. Conventionally, these reactions have been treated statistically assuming energy redistribution in the intermediate complexes. While this is usually valid for large molecules, it is known that S$_N$2 reactions of smaller systems, e.g. halide or hydroxyl anions with halomethanes, show non-statistical behaviour and the full dynamics need to be considered for their appropriate description[4,5,21].

The S$_N$2 reactivity is controlled both by the attacking nucleophile and the atom or molecule that is being substituted, i.e. the leaving group. Generally, a good leaving group is defined by the extent to which it lowers the transition state barrier of the reaction. The ability of a substituent Y to act as a good leaving group is often associated with its basicity or electronegativity, i.e. its ability to accept an additional negative charge[22,23], and the strength of the C-Y bond[10,15]. A direct relation of these parameters to the overall reactivity of an S$_N$2 reaction usually only holds when the leaving groups are structurally similar. If the leaving group abilities of methyl halides are compared, iodine acts as the best and fluorine as the worst leaving group. This order follows the bond strengths of the C-halide bond and the basicity of the leaving halide. The role of the leaving group is often discussed in terms of the overall reactivity of the S$_N$2 reaction, but little is known about how it affects the underlying dynamics.

The combination of crossed-beam scattering and velocity map imaging[24] has provided new insights into S$_N$2 reaction dynamics[9,13]. Combined with direct dynamics simulations this has been successful in explaining gas phase S$_N$2 reaction mechanisms, scattering angle distributions, and energy partitioning among the reaction products[14]. The role of the nucleophile has been the topic of several recent studies. It was found in reactions of $Cl^-$, $F^-$ and $OH^-$ with $CH_3I$, that the nucleophile strongly influences the shape of the entrance channel complex and thus the reaction pathways[9,13,25,26]. In these studies distinct atomic-level mechanisms were found to be important: The *direct rebound mechanism* represents the classical co-linear approach with the product ion leaving in the direction of the incoming reactant ion. The *direct stripping mechanism* attacks the $CH_3$ group from aside and leads to product ions leaving in direction of the incoming neutral reactant. In addition, several *indirect mechanisms*, including the roundabout mechanism[9] were identified, which lead to highly excited products with slow and isotropic ion product distributions.

In the present study we explore the role of the leaving group in bimolecular nucleophilic substitution dynamics by investigating the reaction

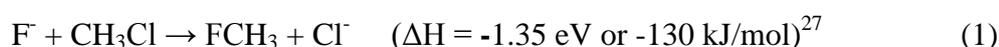
$$F^- + CH_3Cl \rightarrow FCH_3 + Cl^- \quad (\Delta H = -1.35 \text{ eV or } -130 \text{ kJ/mol})^{27} \quad (1)$$

in a combination of experiment and simulation. The stationary atomic configurations of this reaction were adapted from ref. 16 and are displayed in the upper panel of Fig. 1. Reaction (1) is highly exothermic with only a small transition state barrier. A hydrogen-bonded complex of $C_s$ symmetry ($F^-$---$HCH_2Cl$) is found as minimum energy structure in the entrance channel together with a second close lying complex of co-linear $C_{3v}$ symmetry ($F^-$---$CH_3Cl$). The hydrogen-bonded complex had not been included in earlier simulations of the $F^- + CH_3Cl$ reaction[28]. For the reaction

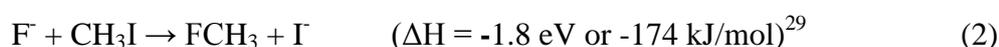
$$F^- + CH_3I \rightarrow FCH_3 + I^- \quad (\Delta H = -1.8 \text{ eV or } -174 \text{ kJ/mol})^{29} \quad (2)$$

very similar entrance channel structures were found[25,26]. The potential energy curve for this reaction was adapted from ref. 26 and is displayed in the lower part of Fig. 1. These structures are in line with a theoretical study on the trends in $S_N2$ reactivity for different halide halomethane combinations that also predicted the hydrogen-bonded complex[10].

Since the branching into direct or indirect reaction dynamics is expected to be influenced mainly by the entrance channel, similar dynamics may be expected for reactions (1) and (2). However, by comparing the results for these reactions - reaction (2) has previously been studied by us in collaboration with Hase and co-workers[25,26] – we find substantial differences that signify the influence of the leaving group on the reaction dynamics, also in the entrance channel.

**Results**

**Differential Scattering Cross-Sections**

In the experimental part of the present study the dynamics of reaction (1) have been examined by measuring differential scattering cross sections with crossed-beam imaging. The experimental setup and procedures are described in the methods section. Three-dimensional probability distributions are obtained for the product velocity vector from between $10^4$ and $10^5$ scattering events. In Fig. 2 (upper left panels) we show the longitudinal and transverse velocity distribution of $Cl^-$ product ions, mapped onto a 2D image for collision energies from 0.6 eV to 2 eV in the centre-of-mass frame for the collision. In this frame the $F^-$ and $CH_3Cl$ reactant velocity vectors align horizontally as shown by the black arrows. For reactive scattering, a kinematical cutoff in velocity can be defined, which defines the highest possible product velocity and is given by the energy available for the reaction. The outermost rings in the images indicate this kinematical cutoff for each collision energy. To guide the eye, also concentric rings are drawn, representing isospheres in translational energy (spaced at 0.5 eV intervals) corresponding to different degrees of internal excitation of the products.

The images for $Cl^-$ in Fig. 2 are dominated by two distinct features, an isotropic feature at nearly zero velocity and one at higher velocities in the backward direction. At the highest collision energy the image shows the $Cl^-$ preferentially scattered into the backward direction. This feature resembles the previously established direct rebound mechanism[9,13], where the nucleophile X attacks the molecule from the $CH_3$ umbrella backside to form a collinear $[X-CH_3-Y]^-$ transient ion. This transient ion dissociates into products with the outgoing $Cl^-$ ion leaving in the same direction as the incoming nucleophile. At the lower collision energies isotropic scattering with slow product velocities is also observed which resembles the indirect complex-mediated mechanism where the available energy is partitioned into various internal degrees of freedom[9]. Similar dynamics are observed for intermediate collision energies for which the relative contributions of both mechanisms shift from indirect to direct dynamics with increasing energy.

The velocity-integrated angular distributions in the centre row of Fig. 2 (black lines) show this change in mechanism in a more quantitative way. The angular distributions evolve from a more forward-backward balanced distribution at low collision energies towards an anisotropic distribution with more events in the backward direction at higher collision energies. These backward scattered $Cl^-$ ions are produced by the direct rebound mechanism. In contrast, direct stripping would lead to mainly forward scattered $Cl^-$ ions[26,30].

From the 2D images we see that the velocity distributions vanish within the kinematical cutoff even at the highest collision energy. Thus, some of the available energy is deposited into rovibrational excitation of the $CH_3F$ product. This is more quantitatively shown in the internal energy distributions in the bottom row of Fig. 2 (black lines), which are extracted from the images. The distributions show a decreasing amount of internal excitation with increasing collision energy, which is another manifestation of the increasing probability for the direct rebound mechanism.

In the right part of Fig. 2 the same type of data are displayed for reaction (2) at 1.53 eV relative collision energy (adapted from ref. 25). A qualitative direct comparison of the scattering image for reaction (2) to those of reaction (1) shows scattering in backward direction and a much larger contribution of slow products. This is quantified both by the angular distribution, which contains a larger isotropic distribution, and by the internal energy distribution, which shows more highly excited product molecules (lower, right panels). Furthermore, in this reaction the overall shape of the scattering images changed only gradually with collision energy[25].

Chemical dynamics simulations have been performed for reaction (1) using the quasi-classical trajectory method on an accurate *ab initio* analytical potential energy surface[16]. Roughly half a million trajectories were run for collision energies in the range of 0.5 to 2.0 eV. Details are provided in the methods section. The computed product internal energy and scattering angle distributions are shown in Fig. 2 (red lines). Essentially all features of the internal energy and angular scattering distributions that are found in the experiment are reproduced in the simulations. This comparison represents an excellent agreement between measurement and simulation for ion-molecule reactions at an unprecedented level of statistics.

**Reaction Mechanisms**

From the experimental and simulated data we have extracted the contribution of the main reaction mechanism, the direct rebound mechanism, to the total scattering signal. This direct rebound fraction is plotted in Fig. 3 (upper panel) as a function of the collision energy. The experimental values are obtained by integrating over the areas marked by the white dashed lines in the images of Fig. 2, normalized to the total counts in each image. The simulation values were derived from the fraction of fast rebound trajectories, defined as reactive trajectories for which the product velocity amounts to more than 1500 m/s and which lead to back-scattered product ions. An analysis of the correlation between the velocity of $Cl^-$ product ion and the integration time revealed that trajectories with final velocities exceeding 1500 m/s also correspond to trajectories with less than 0.58 ps integration time (see inset Fig. 3 lower panel), i.e. which form products in a rapid direct reaction. Slow $Cl^-$ ions are produced by an indirect mechanism, for which the reaction time can be a few or few tens of picoseconds. For the simulated direct rebound fraction very similar results are obtained using either the integration time or the product velocity as selection criterion (see Fig. 3, lower Panel). Given the agreement in the differential scattering distributions, the good agreement for the direct rebound fraction is well expected. The discrepancy at lower relative energies could be explained by an overestimation of the indirect mechanisms in the simulations due to unphysical energy flow during the longer interaction times.

In Fig. 3 the direct rebound mechanism dominates at all but the lowest collision energies for reaction (1) in both experiment (black markers) and simulation (red markers). This observation of a large fraction of direct product ions is in agreement with earlier experimental[31] and theoretical work[28]. Our data also support the interpretation of a guided ion beam study, which showed a large scattering cross section at very low collision energy and identified the hydrogen-bonded complex as explanation for this[27]. At energies above about 0.5 eV a sharp decrease of the cross section was found and assigned to the onset of direct reaction dynamics.

For reaction (2) the observed dynamics are qualitatively different. In this reaction less direct rebound dynamics is found, accompanied up to high collision energies by a lot of internal excitation (see Fig. 3, cyan and blue markers). Experimentally, the direct rebound fraction (cyan markers) is estimated from the data of ref. 25 as marked by the white dashed line in Fig. 2 (right column). The accuracy is determined by varying the lower velocity limit by ±100 m/s. Simulation results using two different levels of theory agree with this smaller direct rebound fraction for $F^-$ + $CH_3I$ (blue full and open markers[20,21]). The role of the direct rebound mechanism is also strongly reduced for the isoelectronic system $OH^-$ + $CH_3I$ (see Fig. 3, green markers[13,30]) which is characterised by an hydrogen-bonded complex as well. Interestingly, the reaction $Cl^-$ + $CH_3I$ features dynamics that bear many similarities with the present reaction[9]. In this reaction (exothermicity 0.55 eV) the minimum energy path is known to follow a collinear geometry with $C_{3v}$ symmetry[9,10]. This suggests that also in the present system the collinear entrance channel is more important for the dynamics than the hydrogen-bonded complex.

To shed light on the origin of the differences for the $Cl^-$ and $I^-$ leaving groups, we have tested the influence of the vibrational coupling between the entrance and the exit channel by simulating the dynamics for a chlorine atom of mass 127 (corresponding to iodine) on the potential obtained for $CH_3Cl$ (Fig. 3 lower panel, blue markers). No significant differences in the rebound fraction are found at high collision energies. At lower collision energies the direct rebound fraction reduces to about half of its value for true Cl, because the cross-section is reduced for the direct channels and enhanced for the indirect channel. Thus, mass-scaling does not account for the observed differences and one has to conclude that the leaving group directly modifies the interaction potential in the entrance channel - enough to strongly change the dynamics, but not enough to change the overall structure and energetics notably.

More insight is gained by inspecting the reaction probability as a function of impact parameter, the opacity function, for each mechanism. Fig. 4 shows this function for the different reaction channels for the two collision energies of 0.55 eV and 1.5 eV for $CH_3Cl$ (upper panel) and 0.3 eV and 1.5 eV for $CH_3I$ (lower panel, from ref. 26). The central panel shows the result for $CH_3Cl'$ ($m_{Cl'}$=127 a.m.u.) at 0.55 eV and 1.5 eV. The total reaction probability (black markers) as well as the individual contributions of indirect (green markers), direct rebound (red markers) and direct stripping mechanism (blue markers) are shown. Comparison of these contributions for the different systems shows that the indirect mechanism extends to similar maximum impact parameters for both reactions (1) and (2). In contrast, the direct rebound mechanism (red markers) is found to extend to larger impact parameters for reaction (1) than for reaction (2) at both energies. The stripping mechanism appears at higher impact parameters and becomes the dominant reaction mechanism at large impact parameters for reaction (2), while it plays only a minor role in reaction (1). The different opacity functions cannot be explained solely by the different mass of the leaving group as an analysis of the impact parameters for $CH_3Cl'$ shows the same range of impact parameters as for reaction (1). Interestingly, both direct reaction mechanisms occur at similar impact parameters at lower and higher collision energy for the reaction with $CH_3Cl$, while with $CH_3I$ the range of impact parameters for which direct rebound or stripping happen strongly depend on the collision energy.

**Discussion**

The occurrence of the direct rebound mechanism at larger impact parameters in the reaction with $CH_3Cl$ suggests the orientation of $CH_3Cl$ by $F^-$ to be more efficient than that of $CH_3I$. A reason might be the larger dipole moment of $CH_3Cl$ ($\mu$ = 1.90 D)[32] compared to $CH_3I$ ($\mu$ = 1.64 D)[32]. The long range interaction of the larger dipole moment with the charge of the $F^-$ anion might allow for a more efficient orientation of the molecule prior to the reaction. This more efficient orientation thus leads to dynamics which follow a direct rebound mechanism even at initially larger impact parameters. Furthermore, a difference in the interaction potential has been found by computing the stability of the frontside attachment of $F^-$ to the halogen atom. The $F^-$---$ICH_3$ complex is bound by 94 kJ/mol while $F^-$---$ClCH_3$ is bound by only 13 kJ/mol. This may enhance the stripping mechanism for $F^-$ + $CH_3I$ and also suppress backside attack for large impact parameters. A correlation of reactant orientation with reactivity in the reaction of $F^-$ with methyl halides was also found by Su *et al.*[28,33]. This correlation was strongest for the reaction of methyl chloride. In their analysis Su *et al.* focused on the overall reactivity of the reaction and energy partitioning. The differential scattering cross-sections obtained in our work allow us to deduce the underlying reaction dynamics.

**Conclusion**

Our results provide insight into the unexpected dynamics and the relative importance of the different reaction mechanisms for $S_N2$ reactions with different leaving groups. The measured energy and angular distributions for the reaction of $F^-$ and $CH_3Cl$ are in excellent agreement with high-level chemical dynamics simulations. From the differential scattering information it has become clear that the direct rebound mechanism dominates at most relative collision energies despite the presence of a hydrogen-bonded entrance channel complex. Different reaction dynamics were found in both experiment and theory for $F^-$ reacting with $CH_3I$ that can not be explained by just changing the mass of the leaving group. Instead we rationalize this with subtle changes in the interaction potential and as a consequence of an increasing range of impact parameters that lead to direct rebound dynamics - likely due to a better orientation of the pre-reaction complex. These results highlight the important role of both nucleophile and leaving group not only for $S_N2$ reactivity but also the underlying reaction dynamics. It will have to be clarified how important this effect is in larger organic reaction systems with even larger dipole moments.

**Methods**

**Experiment**

The experimental setup and data analysis procedure have been published earlier[25,34] and are only summarized here. A pulsed beam of F⁻ ions is produced in a pulsed plasma discharge of 10% $NF_3$ in Ar. The ions are mass-separated by time-of-flight and trapped in an rf octupole ion trap from where they are extracted and crossed with the neutral target molecular beam. **The full width at half maximum of the ion energy distributions** is below typically 150 meV except for the lowest energy, where it is below 250 meV. The pulsed supersonic target beam of $CH_3Cl$, seeded at ~ 5 % in He, is produced in a differentially pumped neutral beam chamber. The translational temperature of the beam in the co-moving frame is ~ 180 K. This moderate cooling was chosen to avoid $CH_3Cl$ clusters. The two beams are made to collide at 60° relative angle in the center of a velocity map imaging spectrometer. Cl⁻ ions are extracted normal to the scattering plane and imaged on a position and time-sensitive detector, which records the 3D velocity vector for each product ion. The relative collision energy is set by tuning the ion energy with an electrostatic decelerator and measuring it with the same spectrometer. From the 3D velocity distributions 2D slice images, energy and angular distributions are obtained by numerical integration. For visualization the 2D images are gently smoothed.

**Simulation**

Quasi-classical trajectory (QCT) computations are performed for the ground-state F⁻ + $CH_3Cl(v = 0)$ reaction using a global *ab initio* full-dimensional potential energy surface (PES)[16]. The analytical PES was obtained by fitting all-electron (AE) CCSD(T)/aug-cc-pCVQZ-quality composite ab initio energy points. The stationary points of the PES were characterized by the focal point analysis (FPA) method considering: (a) extrapolation to the complete basis set limit using the AE-CCSD(T)/aug-cc-pCVnZ [n = Q(4) and 5] energies, (b) post-CCSD(T) correlation effects up to CCSDT(Q) using DZ-quality bases, and (c) scalar relativistic effects at the Douglas–Kroll AE-CCSD(T)/aug-cc-pCVQZ level of theory[16,35]. The interaction potential for frontside attachment was computed at the CCSD(T)-F12b/aug-cc-pVDZ(-PP) level of theory.

Standard normal mode sampling is used to prepare the vibrational ground state ($v = 0$) of $CH_3Cl$ and the rotational angular momentum is set to zero. The initial orientation of $CH_3Cl$ is randomly sampled and the distance between F⁻ and $CH_3Cl$ is $(x^2 + b^2)^{1/2}$, where $b$ is the impact parameter, scanned from 0 to $b_{max}$ with a step size of 0.5 bohr and $x$ is set to 20 bohr. We have run trajectories at collision energies of 0.55, 0.9, 1.2, 1.5, and 2.0 eV with $b_{max}$

values 9.0, 8.0, 7.0, 7.0, and 7.0 bohr and with 95000, 85000, 75000, 75000, and 75000 trajectories, respectively. The integration time step is 0.0726 fs and each trajectory is propagated until the maximum of the actual inter-atomic distances is 1 bohr larger than the initial one. We have found that basically no trajectory violates the product zero-point energy; thus, the QCT analysis considers all the trajectories. We have also performed QCT computations for the mass-scaled reaction $F^- + CH_3Cl'$ (v = 0) by setting the mass of Cl' to 127 a.m.u. and using the PES of $F^- + CH_3Cl$. Here slightly larger $b_{max}$ values of 12.0, 9.5, 8.0, 8.0, and 8.0 bohr are used resulting in 125000, 100000, 85000, 85000, and 85000 trajectories at collision energies of 0.55, 0.9, 1.2, 1.5, and 2.0 eV, respectively. All the other computational details are the same as described above.

**References**

1. Vollhardt, K. P. C. & Shore, N. E. *Organic Chemistry, Structure and Function*. (Pallgrave Macmillan, 2007).

2. Olmstead, W. N. & Brauman, J. I. Gas-phase nucleophilic displacement reactions. *J. Am. Chem. Soc.* **99,** 4219–4228 (1977).

3. Shaik, S. S. The collage of $S_N2$ reactivity patterns - a state correlation diagram model. *Prog. Phys. Org. Chem.* **15,** 197–337 (1985).

4. Viggiano, A. A., Morris, R. A., Paschkewitz, J. S. & Paulson, J. F. Kinetics of the gas-phase reactions of chloride anion, $Cl^-$ with $CH_3Br$ and $CD_3Br$: experimental evidence for nonstatistical behavior? *J. Am. Chem. Soc.* **114,** 10477–10482 (1992).

5. Hase, W. L. Simulations of gas-phase chemical reactions: applications to $S_N2$ nucleophilic substitution. *Science* **266,** 998–1002 (1994).

6. Chabinyc, M. L., Craig, S. L., Regan, C. K. & Brauman, J. I. Gas-phase ionic reactions: dynamics and mechanism of nucleophilic displacements. *Science* **279,** 1882–1886 (1998).

7. Laerdahl, J. K. & Uggerud, E. Gas phase nucleophilic substitution. *Int. J. Mass Spectrom.* **214,** 277–314 (2002).

8. Schmatz, S. Quantum dynamics of gas-phase $S_N2$ reactions. *ChemPhysChem* **5,** 600–617 (2004).

9. Mikosch, J. *et al.* Imaging nucleophilic substitution dynamics. *Science* **319,** 183–6 (2008).

10. Bento, A. P. & Bickelhaupt, F. M. Nucleophilicity and leaving-group ability in frontside and backside $S_N2$ reactions. *J. Org. Chem.* **73,** 7290–9 (2008).

11. Garver, J. M., Gronert, S. & Bierbaum, V. M. Experimental validation of the alpha-effect in the gas phase. *J. Am. Chem. Soc.* **133,** 13894–13897 (2011).

12. Kretschmer, R., Schlangen, M. & Schwarz, H. Efficient and selective gas-phase monomethylation versus N-H bond activation of ammonia by bare $Zn(CH_3)^+$: atomic zinc as a leaving group in an $S_N2$ reaction. *Angew. Chemie (International Ed.)* **50,** 5387–5391 (2011).

13. Otto, R. *et al.* Single solvent molecules can affect the dynamics of substitution reactions. *Nature Chem.* **4,** 534–8 (2012).

14. Xie, J. *et al.* Identification of atomic-level mechanisms for gas-phase $X^- + CH_3Y$ $S_N2$ reactions by combined experiments and simulations. *Acc. Chem. Res.* **47,** 2960–9 (2014).

15. Fernández, I. & Bickelhaupt, F. M. The activation strain model and molecular orbital theory: understanding and designing chemical reactions. *Chem. Soc. Rev.* **43,** 4953 (2014).

16. Szabó, I. & Czakó, G. Revealing a double-inversion mechanism for the F⁻+CH$_3$Cl S$_N$2 reaction. *Nat. Commun.* **6,** 5972 (2015).

17. Thallmair, S., Kowalewski, M., Zauleck, J. P. P., Roos, M. K. & de Vivie-Riedle, R. Quantum dynamics of a photochemical bond cleavage influenced by the solvent environment: A dynamic continuum approach. *J. Phys. Chem. Lett.* **5,** 3480–3485 (2014).

18. Orr-Ewing, A. J. Perspective: Bimolecular chemical reaction dynamics in liquids. *J. Chem. Phys.* **140,** 090901 (2014).

19. Garver, J. M. *et al.* A direct comparison of reactivity and mechanism in the gas phase and in solution. *J. Am. Chem. Soc.* **132,** 3808–3814 (2010).

20. Liu, S., Hu, H. & Pedersen, L. G. Steric, quantum, and electrostatic effects on S$_N$2 reaction barriers in gas phase. *J. Phys. Chem. A* **114,** 5913–5918 (2010).

21. DeTuri, V. F., Hintz, P. A. & Ervin, K. M. Translational activation of the S$_N$2 nucleophilic displacement reactions Cl⁻ + CH$_3$Cl (CD$_3$Cl) → ClCH$_3$ (ClCD$_3$) + Cl⁻ : A guided ion beam study. *J. Phys. Chem. A* **101,** 5969–5986 (1997).

22. Anderson, J. S. M., Liu, Y., Thomson, J. W. & Ayers, P. W. Predicting the quality of leaving groups in organic chemistry: Tests against experimental data. *J. Mol. Struct. THEOCHEM* **943,** 168–177 (2010).

23. Jaramillo, P., Domingo, L. R. & Pérez, P. Towards an intrinsic nucleofugality scale: The leaving group (LG) ability in CH$_3$LG model system. *Chem. Phys. Lett.* **420,** 95–99 (2006).

24. Eppink, A. T. J. B. & Parker, D. H. Velocity map imaging of ions and electrons using electrostatic lenses: Application in photoelectron and photofragment ion imaging of molecular oxygen. *Rev. Sci. Instrum.* **68,** 3477 (1997).

25. Mikosch, J. *et al.* Indirect dynamics in a highly exoergic substitution reaction. *J. Am. Chem. Soc.* **135,** 4250–9 (2013).

26. Sun, R., Davda, C. J., Zhang, J. & Hase, W. L. Comparison of direct dynamics simulations with different electronic structure methods. F⁻ + CH$_3$I with MP2 and DFT/B97-1. *Phys. Chem. Chem. Phys.* **17,** 2589–97 (2015).

27. Angel, L. A. & Ervin, K. M. Dynamics of the gas-phase reactions of fluoride ions with chloromethane. *J. Phys. Chem. A* **105,** 4042–4051 (2001).

28. Su, T., Wang, H. & Hase, W. L. Trajectory studies of S$_N$2 nucleophilic substitution F⁻ + CH$_3$Cl → FCH$_3$ + Cl⁻. *J. Phys. Chem. A* **102,** 9819–9828 (1998).

29. Zhang, J. *et al.* F⁻ + CH$_3$I → FCH$_3$ + I⁻ Reaction dynamics. nontraditional atomistic mechanisms and formation of a hydrogen-bonded complex. *J. Phys. Chem. Lett.* **1,** 2747–2752 (2010).

30. Xie, J. *et al.* Direct dynamics simulations of the product channels and atomistic mechanisms for the OH⁻ + CH$_3$I reaction. Comparison with experiment. *J. Phys. Chem. A* **117,** 7162–7178 (2013).

31. Vanorden, S. L., Pope, R. M. & Buckner, S. W. Energy disposal in gas-phase nucleophilic displacement reactions. *Org. Mass Spectrom.* **26,** 1003–1007 (1991).

32. *Handbook of Chemistry and Physics*. (CRC Press, 2003).

33. Su, T., Morris, R. A., Viggiano, A. A. & Paulson, J. F. Kinetic energy and temperature dependences for the reactions of fluoride with halogenated methanes: experiment and theory. *J. Phys. Chem.* **94,** 8426–8430 (1990).

34. Wester, R. Velocity map imaging of ion-molecule reactions. *Phys. Chem. Chem. Phys.* **16,** 396–405 (2014).

35. Szabó, I., Császár, A. G. & Czakó, G. Dynamics of the F⁻ + CH$_3$Cl → Cl⁻ + CH$_3$F S$_N$2 reaction on a chemically accurate potential energy surface. *Chem. Sci.* **4,** 4362 (2013).

**Acknowledgements**

This work is supported by the Austrian Science Fund (FWF), project P 25956-N20. E.C. acknowledges support from a DOC-Fellowship by the Austrian Academy of Sciences ÖAW. G.C. was supported by the Scientific Research Fund of Hungary (OTKA, PD-111900) and the János Bolyai Research Scholarship of the Hungarian Academy of Sciences.

**Author contributions**

M.S., E.C., A.K., and R.W. conceived the experiment. M.S., E.C., A.K., and M.K performed the measurements. M.S. analysed the data. I.S. and G.C. carried out the simulations. All authors discussed the results. A.K, M.S., E.C., J.M., G.C. and R.W. wrote the paper.

**Competing Interests**

The authors declare that they have no competing financial interests.

**Correspondence**

Correspondence should be addressed to Roland Wester (roland.wester@uibk.ac.at) or Gábor Czakó (email: gczako@chem.u-szeged.hu)

**Figure 1: Calculated minimum energy path of reactions (1) and (2) along the reaction pathway.** Stationary points along the reaction pathway are shown for the reaction of **(a)** F$^-$ + CH$_3$Cl obtained by a relativistic all-electron CCSDT(Q)/complete-basis-set-quality composite method (FPA Method, see methods section, adapted from ref. 16) and **(b)** F$^-$ + CH$_3$I (adapted from ref. 26). Both systems show the same qualitative behavior. The structures of the stationary points are illustrated in the centre. Two pre-reaction complexes exist: the energetically lower H-bonded complex F$^-$---HCH$_2$X and a second one of $C_{3v}$ symmetry F$^-$---CH$_3$X.

**Figure 2: Differential scattering cross sections and extracted angular and energy distributions of the S$_N$2 reactions F$^-$ + CH$_3$Cl and F$^-$ + CH$_3$I**

The upper left panels show the velocity distribution of Cl$^-$ product ions in the center-of-mass frame at five different relative collision energies (0.6 - 2 eV). The concentric rings mark spheres of equal product internal energies spaced at 0.5 eV with the outermost solid ring indicating the kinematical cut-off. The Newton diagram at the top illustrates the relative orientation of the velocity vectors of the reactants and the Cl$^-$ product ions. Several dynamical features can be clearly distinguished in the images: an isotropic feature at nearly zero velocity, one backscattered distribution at velocity and a small forward scattered contribution for high collision energies. The white dashed lines indicate the area used to determine the direct rebound fraction. The central panel displays the velocity integrated angular distributions for Cl$^-$ ions (exp.: black, sim.: red). The chemical dynamics simulations compare well with the experiment. The lowest panel shows the internal energy distribution of the CH$_3$F product molecules (exp.: black, sim.: red). The outermost right column shows previous results for the reaction of F$^-$ + CH$_3$I at 1.5 eV relative collision energy (adapted from ref. 25). Here dynamics are visible: a larger fraction of slow product ions, and therefore a higher degree of internal excitation of CH$_3$F products, is found together with a larger the degree of isotropic scattering.

**Figure 3: Direct rebound fraction for F$^-$ + CH$_3$Cl in comparison to previously studied systems**

*Both panels:* The experimental normalized yield of the direct rebound fraction obtained from the white dashed area in the images in Figure 1 is displayed as a function of relative energy (black circles). The given errorbars reflect the counting statistics. The experimental values compare well with the simulation results (red squares) obtained with the same angular and velocity cuts. *Upper panel:* Additionally, values for the reaction of $F^-$ with $CH_3I$ are included in blue. Cyan circles indicate exp. data obtained by integration from ref. 25. The solid line connects the upper and lower limits. Simulation results from ref. 25 and 26 are shown as closed and open blue squares. Reaction (1) shows a much higher direct rebound fraction than Reaction (2). This deviation of rebound fraction increases with increasing collision energy. For comparison, data for the reaction $OH^- + CH_3I$ is included in green color (circles: exp., squares: sim.)[13,30]. $OH^-$ is isoelectronic to $F^-$ and the behavior of the system resembles that of reaction (2). *Lower panel:* Direct rebound fraction from experiment and simulation. A fraction obtained from fast trajectories with t < 0.58 ps (green squares) matches the direct rebound fraction obtained from the velocity cut v > 1500 m/s. This can be also inferred from the inset which shows the correlation of ion product speed and collision time on a logarithmic scale: the fastest $Cl^-$ ions are produced in the shortest time, via the direct rebound mechanism. The direct rebound fraction for the reactant $CH_3Cl'$ ($m_{Cl'}$ = 127 a.m.u.; blue squares) is slightly lower than for $CH_3Cl$ but this change of mass does not explain the enhanced direct rebound fraction of reaction (1) compared to reaction (2).

**Figure 4: Opacity functions for different reaction mechanisms**
The reaction probability is shown as a function of the impact parameter. The total probability (black circles) is split into the individual contributions of indirect (green circles), direct rebound (red circles) and stripping (blue circles) mechanism. In the two upper rows the left column shows the results for 0.55 eV relative collision energy and the right one for 1.5 eV. $F^-$ + $CH_3Cl$ is displayed in the top panel, $F^-$ + $CH_3Cl'$ ($m_{Cl'}$ = 127 a.m.u.) in the middle panel. In bottom panel for the reaction $F^-$ + $CH_3I$ data from ref. 26 at a relative energy of 0.3 eV (left) and 1.5 eV (right) is shown. The results for $CH_3Cl$ and $CH_3Cl'$ are qualitatively the same but both deviate from those of $CH_3I$. The direct rebound mechanism occurs at a larger range of impact parameters for reaction (1) and shows less energy dependence. Also stripping plays a much larger role for reaction (2).

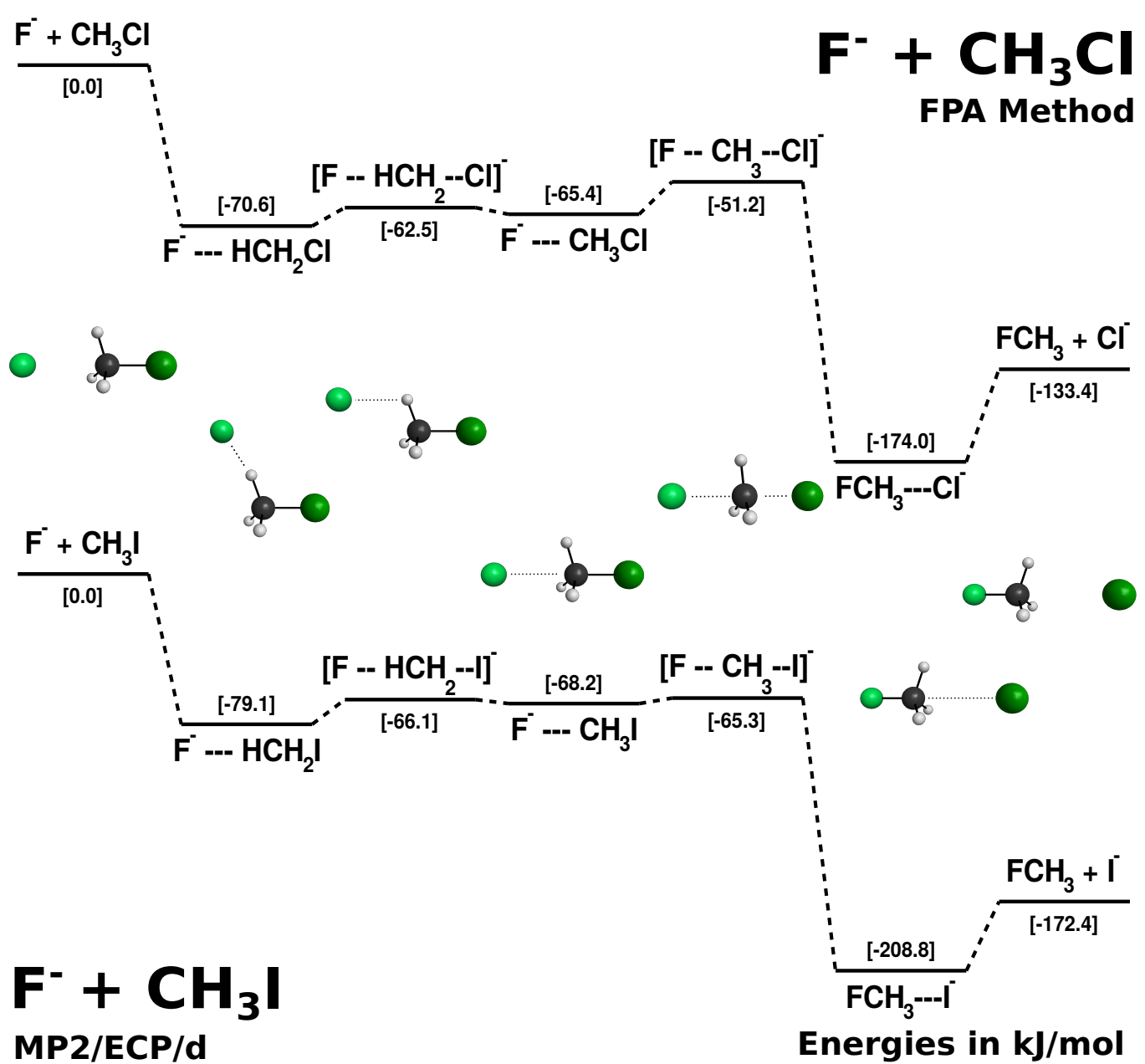

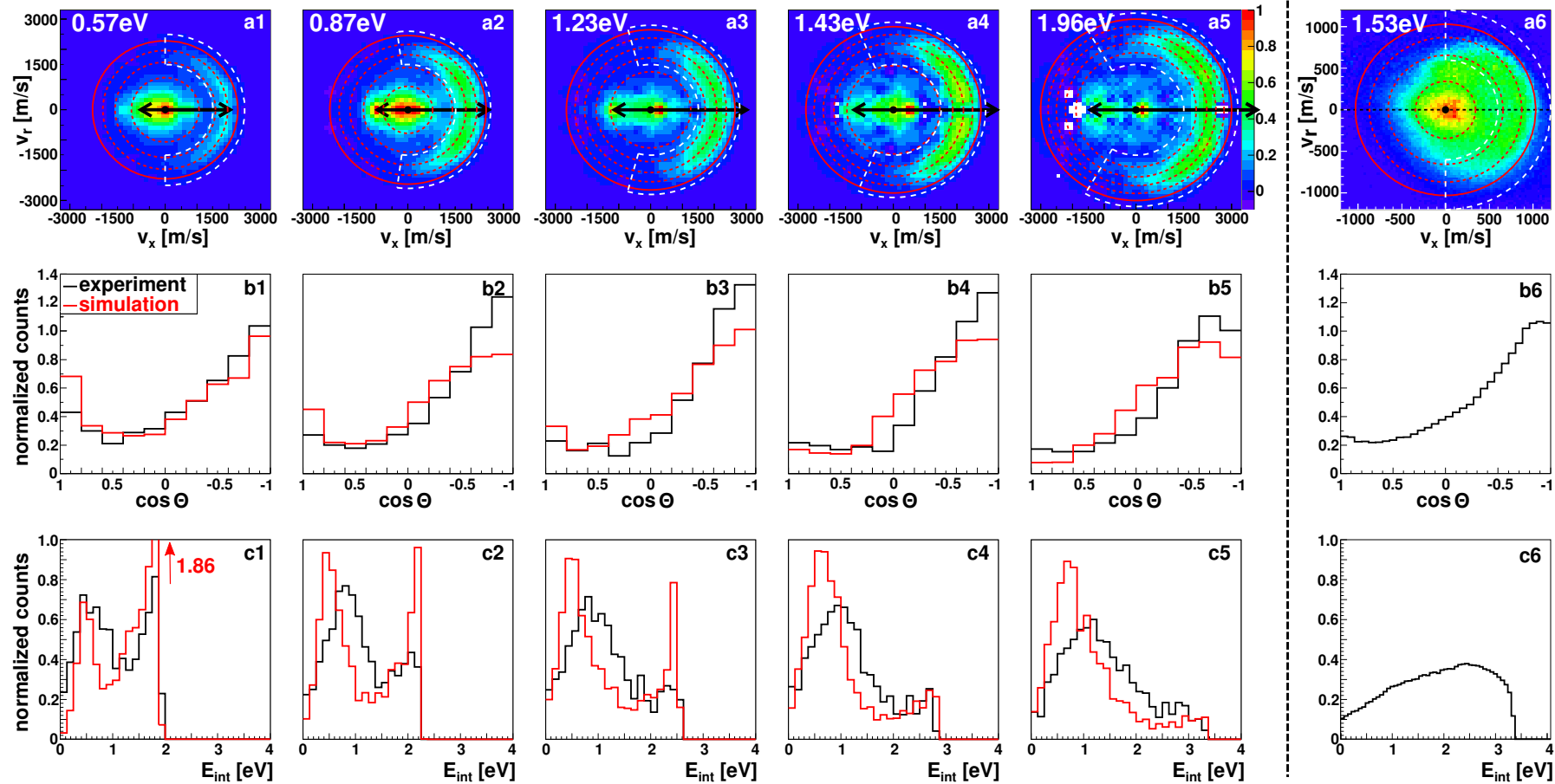

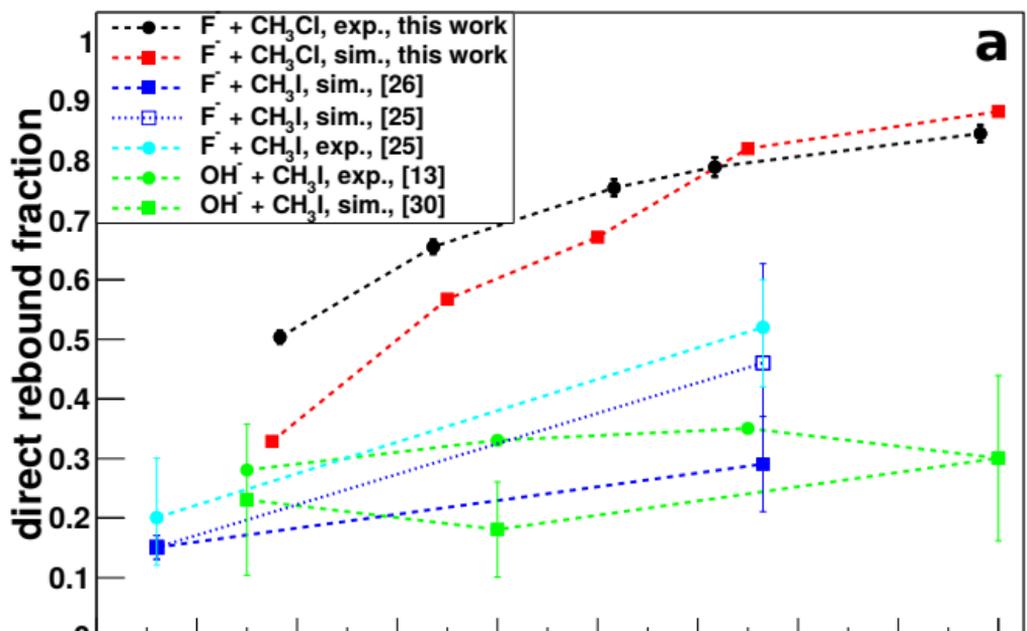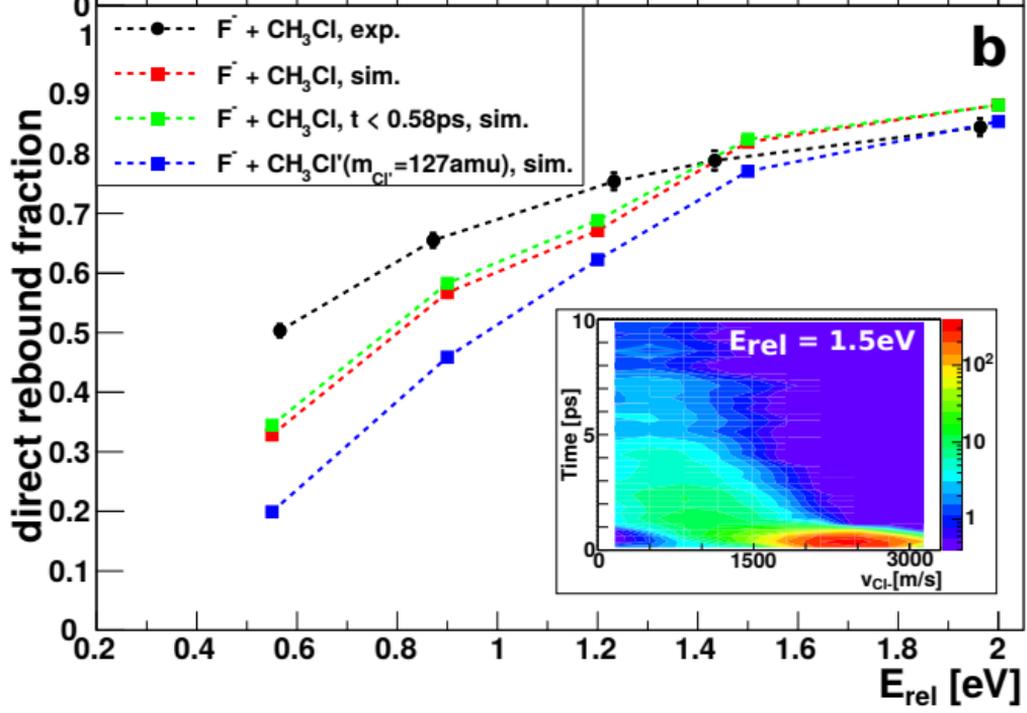

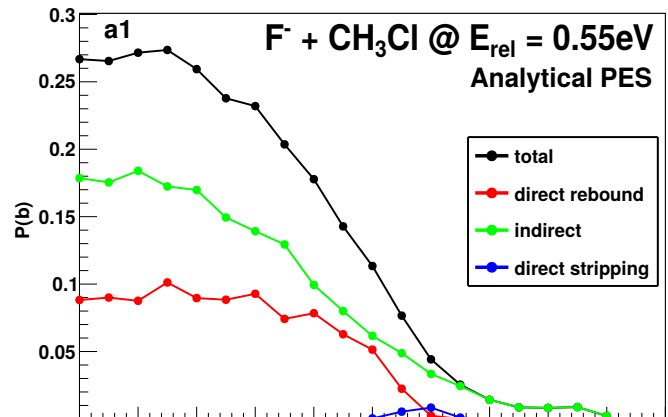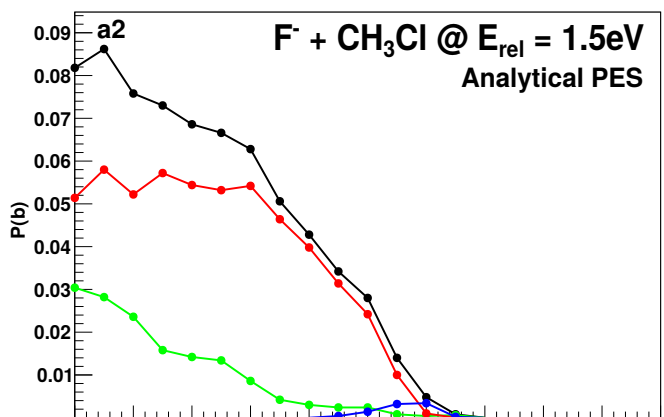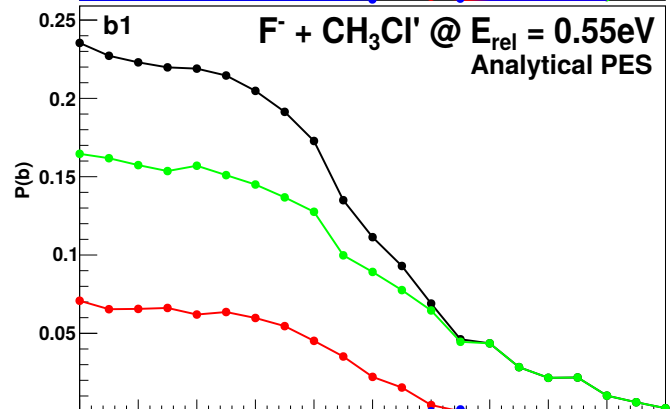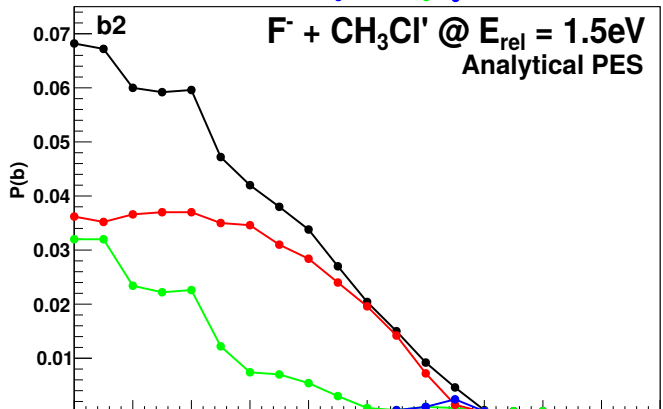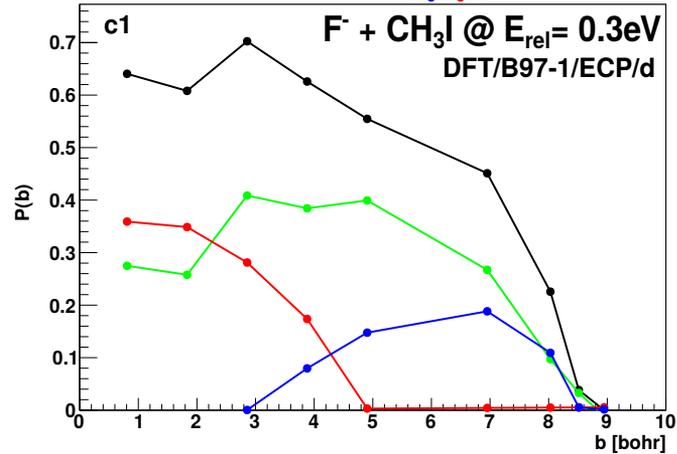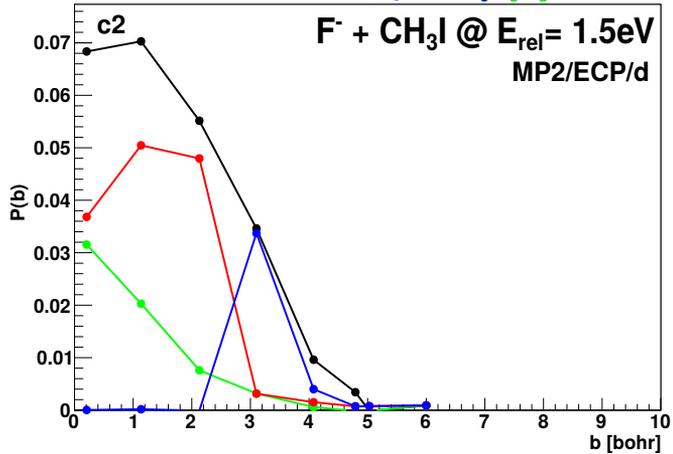